\documentclass[aps,preprint,floatfix,nofootinbib,showpacs]{revtex4}
\usepackage{graphicx,color,epstopdf,amsmath}

\begin{document}
\title{Top Quark Forward-Backward Asymmetry \\
in the Large Invariant Mass Region}

\renewcommand{\thefootnote}{\fnsymbol{footnote}}

\author{ 
Kingman Cheung$^{1,2,3}$, 
and Tzu-Chiang Yuan$^4$
 }
\affiliation{
$^1$Division of Quantum Phases \& Devices, School of Physics, 
Konkuk University, Seoul 143-701, Republic of Korea \\
$^2$Department of Physics, National Tsing Hua University, 
Hsinchu 300, Taiwan
\\
$^3$Physics Division, National Center for Theoretical Sciences,
Hsinchu 300, Taiwan
\\
$^4$Institute of Physics, Academia Sinica, Nankang, Taipei 11529, Taiwan
}

\renewcommand{\thefootnote}{\arabic{footnote}}
\date{\today}

\begin{abstract}
  The forward-backward asymmetry (FBA) in top-pair production that was
  observed in 2008 gets a boost in a recent CDF publication.  Not only
  has the FBA further been confirmed, but also distributional
  preferences are shown.  Strikingly, the FBA is the most sizable in
  the large $M_{t\bar t}$ invariant mass region and in the large
  rapidity difference $|\Delta y|$ region.  Here we used our
  previously proposed $t$-channel exchanged $W'$ boson to explain the
  new observations. We show that a new particle exchanged in the
  $t$-channel generically gives rise to such observations.
  Furthermore, we show that the proposed $W'$ can be directly produced
  in association with a top quark at the Tevatron and the LHC.  We
  perform a signal-background analysis and show that such a $W'$ is
  readily observed at the Tevatron with a 10 fb$^{-1}$ luminosity and
  at the LHC-7 with just a 100 pb$^{-1}$ luminosity.
\end{abstract}


\maketitle
 
\section{Introduction} 
The top quark was the last piece of quarks that was discovered more than
15 years ago \cite{top1,top2}. While waiting for the Higgs boson at
the LHC, the top quark has been making some noise about the presence of 
new physics. The forward-backward 
asymmetry (FBA) in top-quark pair production was found in 2008
by CDF  \cite{cdf-old} and by D\O \cite{D0-old}.
While the Standard Model (SM) only predicts a level 
as small as a few percent arising from the higher-loop contributions, 
the measurement by CDF \cite{cdf-old}, however, was as large as
\begin{equation}
\label{asy}
  A^{t\bar t} \equiv \frac{N_t( \cos\theta >0) - N_t(\cos\theta < 0)}
 {N_t( \cos\theta >0) + N_t(\cos\theta < 0)} = 0.19 \pm 0.065\;({\rm stat})
   \pm 0.024\; ({\rm syst})  \;,
\end{equation}
where $\theta$ is the production angle of the top quark $t$ 
in the $t \bar t$ rest frame.  The measurement in the $p\bar p$ laboratory frame
is correspondingly smaller, because of the Lorentz boost (washout) 
of the partons along the beam axis.  

The anomaly did not die out, but gets a reconfirmation in a recent CDF
publication \cite{cdf1}.  With a larger data set ($5.3$ fb$^{-1}$) the
FBA persists at the level of $A^{t\bar t} = 0.158 \pm 0.074$ \cite{cdf1},
which is a less-than-$2\sigma$ effect
 after subtracting the SM contribution of 
$A^{t\bar t}_{\rm SM} = 0.058 \pm 0.009$ \cite{mcfm,asymm}. 
Though the deviation is slightly
smaller, the most striking feature is that the FBA shows
distributional preferences.
The FBA is the most obvious in the large $M_{t\bar t}$ 
invariant mass region 
and in the large rapidity difference $|\Delta y|$ region.
The analysis in the CDF paper \cite{cdf1}
showed that the FBA is consistent with zero for $M_{t\bar t} < 450$ GeV
but a larger-than-$3\sigma$
effect in $M_{t\bar t} > 450$ GeV region.
At the same time, the analysis also showed that the 
FBA is large in the large rapidity difference $\Delta y$ region, 
which is a $2\sigma$ 
effect.  They are summarized in the third last row in Table~\ref{table1},
where we also show in the second last row the SM predictions from the MCFM
\cite{mcfm}.

If the FBA is true, it will indicate the presence of new physics, 
because within the SM the asymmetry is only up to about 5\% \cite{asymm}.  
In the past two years, numerous works have been carried out to explain
the anomaly 
\cite{1,2,3,4,5,6,7,8,9,10,11,12,13,ours,14,15,16,17,18,Barger:2010mw}.
The explanations can be divided into two categories: (i) a $t$-channel
exchanged particle such as a $W'$ or a $Z'$ with flavor-changing couplings
between the top quark and the $d$ or $u$ quark, and (ii) a heavy $s$-channel
exchanged particle such as an axial-gluon with specific couplings to
the top quark and the light quarks.  In the latter case,
the couplings are somewhat contrived
in order to achieve a positive FBA.  

In a previous work \cite{ours}, we proposed an extra $W'$ boson 
that only couples to the $d$ and $t$ quarks.  Thus, the $d\bar d$ 
initial state turns into the $t\bar t$ final state via a charged-current 
exchange of the $W'$ boson in the $t$-channel.  We mapped out the suitable
parameter space of the $W'$ mass $M_{W'}$ and the coupling $g'$.
In this Letter, we show that such a $t$-channel exchanged particle can
easily accommodate a FBA and also that it naturally gives rise to a
large FBA in the large $M_{t\bar t}$ region and in the large $|\Delta
y|$ region. We show that for $M_{W'} = 200 - 600$ GeV with appropriate
couplings we can bring the predictions to be within
$1-1.5\,\sigma$ of the data.
This is our main result.

In addition, since the $W'^\pm$ boson proposed is relatively light, only 
$200-600$ GeV, it can be directly produced in association with a 
top quark or antiquark at the Tevatron and the LHC. The $W'^-\;(W'^+)$ 
produced would then decay right away into $\bar t d\; (t\bar d)$, 
giving rise to a top-quark pair plus one jet in the final state.
The irreducible background is QCD production of $t\bar t+1j$.  
We show that the increase in $t\bar t$ production by direct $W'$
production is within $1\sigma$ error of the measured $t\bar t$ 
cross section.  We  perform a signal-background analysis based on
parton-level calculations. The cleanest signal of $W'$ production
would be the sharp peak of the invariant mass $M_{tj}$ distribution. 
We require one top quark to decay hadronically while the other one
semi-leptonically. In this case, one has less confusion in jet combinations,
and one can still fully reconstruct the hadronic top 
and combine with the light jet to form the peak of $W'$.  We expect
the background to give a continuum in the $M_{tj}$ distribution.  Thus,
we can count the number of events below the peak for the signal and
background.  At the end, we can see that the Tevatron can observe such
a $W'$ up to 400 GeV while the LHC operating at 7 TeV can observe
such a $W'$ almost immediately.  This is an important result of this work.

Improvements over our previous work are as follows.
\begin{enumerate}
\item
 We use the top-quark mass $m_t = 172.5$ GeV and the most recent
published $t\bar t$ cross section,  which are the same as in 
the most recent CDF publications \cite{cdf1,cdf2}.

\item
 We calculate the FBA as functions of $|\Delta y|$ and $M_{t\bar t}$.
Through the figures it is clear that larger FBA in the large
$|\Delta y|$ and large $M_{t\bar t}$ region is a generic feature of
a new $t$-channel exchanged particle.

\item
With the additional charged $W'$ boson that we proposed, we can bring 
the overall $A^{t\bar t}$ to be around the measured value, the 
$A^{t\bar t} (|\Delta y|>1)$ within $1\sigma$,
and $A^{t\bar t} (M_{t\bar t}> 450 \;{\rm GeV} )$ within $1.5\sigma$.

\item 
We calculate the direct production of the $W'$ associated with a top
quark at the Tevatron and the LHC.  We show that in the presence of
irreducible background of $t\bar t j$ the $W'$ up to about 400 GeV
could be observed at the Tevatron.  On the hand, the $W'$ all the way
to 600 GeV could be easily observed at the LHC. 

\end{enumerate}

\section{The Forward-Backward Asymmetry}
The production angle $\theta$ in the $t\bar t$ rest frame is related to
the rapidity of the $t$ and $\bar t$ in the $p\bar p$ frame by
\begin{equation}
\label{yt-costheta}
 \Delta y \equiv
  y_t - y_{\bar t} = 2 \; {\rm arctanh} \; \left( \sqrt{ 
  1 - \frac{4 m^2_t}{\hat s} }  \; \cos\theta \right )
\end{equation}
where $\hat s$ is the square of the center-of-mass energy of the $t\bar t$ 
pair.  Therefore, the difference $\Delta y$ 
between the rapidities of the $t$ and
$\bar t$ in the $p\bar p$ frame is a close measure of the production angle
in the $t\bar t$ frame.  Moreover, the sign of $\Delta y$ is the
same as $\cos\theta$, such that the asymmetry in Eq. (\ref{asy}) can be
given by
\begin{equation}
\label{asy-yt}
 A^{t\bar t} \equiv \frac{N_t( \Delta y >0) - N_t( \Delta y < 0)}
 {N_t( \Delta y >0) + N_t( \Delta y < 0)}   \;.
\end{equation}
Our parton level calculation uses this definition to calculate
the FBA.

Suppose the interaction vertex for the $W'$
boson with the down and top quarks is given by
\begin{equation}
 {\cal L} = - g' \; W^{'+}_\mu \; \bar t \gamma^\mu \left( g_L P_L + g_R P_R 
\right ) d   \ + \hbox{ h.c.} \;  \;,
\end{equation}
where $P_{L,R} = (1 \mp \gamma^5)/2$ are the chirality projection operators,
$g_{L,R}$ are the chiral couplings of the $W'$ boson with fermions, and 
$g'$ is the coupling constant. In Ref.~\cite{ours}, we demonstrated 
that the pure right-handed 
coupling where $g_L=0$ and $g_R=1$ can fit the data in a more consistent way.
Also, the pure right-handed $W'$ is less constrained by 
the $SU(2)_L$ symmetry.
We therefore focus on this case of pure right-handed coupling in what follows.

The process $ d (p_1) \; \bar d (p_2) \to t (k_1) \; \bar t(k_2)$ is 
described by two Feynman diagrams, 
one $s$-channel diagram from the one gluon exchange
and one $t$-channel diagram from the $W'$ exchange. Ignoring the $d$ quark mass,
the spin- and color-summed amplitude squared is given by
\begin{eqnarray}
\sum \left |{\cal M} \right |^2  &=& \frac{9 g'^4}{t_{W'}^2} \biggr[ 
     4 \left( (g_L^4 + g_R^4) u_t^2 
   + 2 g_L^2 g_R^2 \hat s ( \hat s - 2 m_t^2 ) \right )
+ \frac{ m_t^4}{m_{W'}^4 } 
     ( g_L^2 + g_R^2 )^2 ( t_t^2 + 4 m_{W'}^2 \hat s )      \biggr ] 
\label{5}
\\
&+& \frac{16 g_s^4}{\hat s^2} \, \left( u_t^2 + t_t^2 + 2 \hat s m_t^2 \right )
 +  \frac{16 g'^2 g_s^2}{ \hat s \, t_{W'} } ( g_L^2 + g_R^2 ) \,
         \left [ 2 u_t^2 + 2 \hat s m_t^2 
      + \frac{m_t^2}{m_{W'}^2 }( t_t^2 + \hat s m_t^2 ) \right ]
\ ,
  \nonumber 
\end{eqnarray}
where 
$\hat s = (p_1 + p_2)^2$, $t = (p_1 - k_1)^2$, $u = (p_1 - k_2)^2$ 
and
\begin{equation}
       t_t = t - m_t^2 =-\hbox{$1\over2$}{\hat s}(1-\beta\cos\theta)
\,,\;\;\;  u_t = u - m_t^2 =-\hbox{$1\over2$}{\hat s}(1+\beta\cos\theta) 
\,, \;\;\; t_{W'} = t - m_{W'}^2\;,
\end{equation}
with $\beta = \sqrt{ 1 - 4 m_t^2/ \hat s}$. 
The initial spin- and color-averaged amplitude squared is given by
\begin{equation}
  \overline{\sum \left| {\cal M} \right |^2 }  
  = \frac{1}{4} \, \frac{1}{9} \; {\sum \left| {\cal M} \right |^2 }   \;.
\end{equation}
The differential cross section versus the cosine of the production angle 
$\theta$ is 
\begin{equation}
 \frac{ d \hat \sigma}{ d \cos\theta} = \frac{\beta} {32 \pi \hat s}\,
  \overline{\sum \left| {\cal M} \right |^2 }  \;,
\end{equation}
where $\hat \sigma$ denotes the cross section for the subprocess which
is then folded with the parton distribution functions to obtain the
measured cross section.
The FBA is obtained by integrating over the positive and 
negative range of the $\cos\theta$ variable.

\begin{figure}[th!]
\centering
\includegraphics[width=5in]{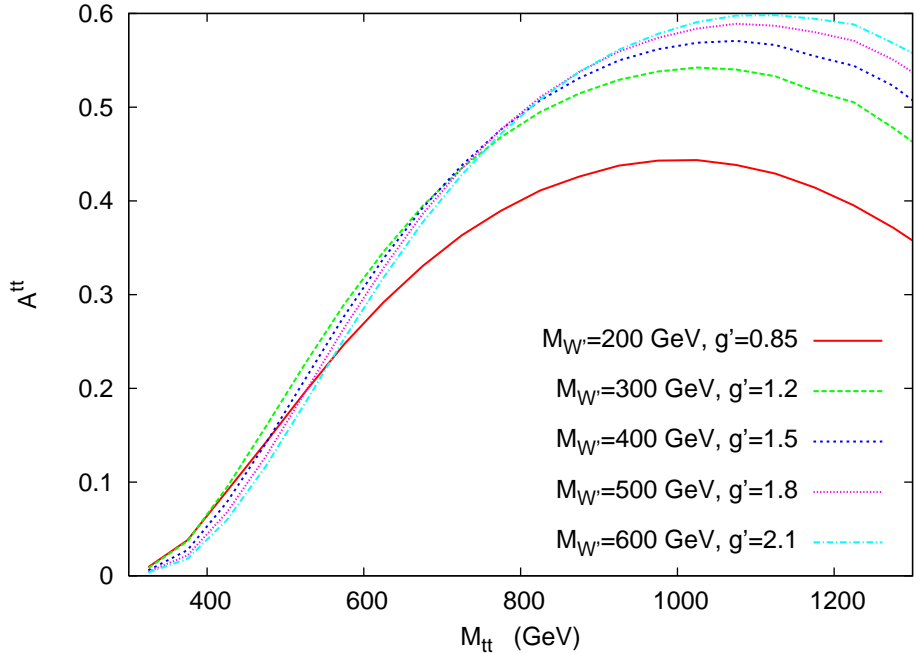}
\caption{\small \label{mtt}
The forward-backward asymmetry of top-pair production versus
the invariant mass $M_{t\bar t}$ at the Tevatron for
various values of $M_{W'}$ and $g'$.}
\end{figure}

\begin{figure}[th!]
\centering
\includegraphics[width=5in]{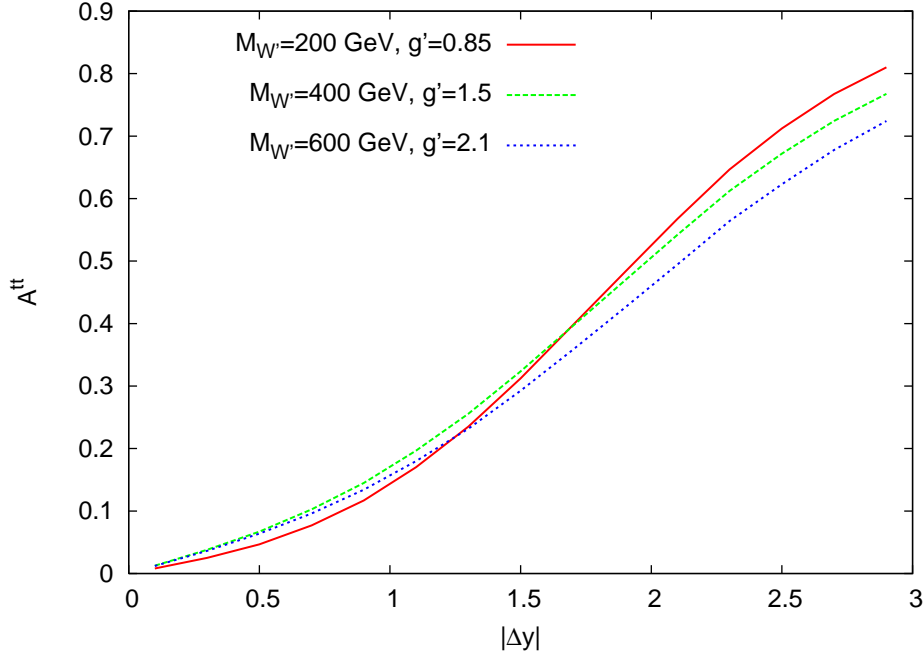}
\caption{\small \label{dy}
The forward-backward asymmetry of top-pair production versus
$|\Delta y| \equiv |y_t - y_{\bar t}|$ at the Tevatron for
various values of $M_{W'}$ and $g'$.}
\end{figure}

We can also easily calculate the invariant mass $M_{t\bar t}$ 
distribution in the forward and backward directions, through which 
we can calculate the FBA versus the invariant mass.  We show the
FBA versus $M_{t\bar t}$ for various values of $M_{W'}$ and $g'$ in 
Fig.~\ref{mtt}.  The values
are chosen such that it can bring the predictions within $1-1.5 \sigma$ 
of the data without violating the constraints on total cross sections
and invariant mass distribution \cite{ours}.
We also use the $\Delta y$ distribution, in which the forward direction 
$(\Delta y >0)$ has more events than the backward direction $(\Delta y <0)$,
to calculate the FBA versus $|\Delta y|$, as shown in 
Fig.~\ref{dy}.  It is clear from Figs.~\ref{mtt} and \ref{dy} that
the FBA becomes large in the large $M_{t\bar t}$ region and in the
large $|\Delta y|$ region.  This is a generic feature for a new particle
exchanged in the $t$-channel, whether it is a $W'$, $Z'$, or a scalar boson.

\begin{table}[b!]
\caption{\small The forward-backward asymmetry for top-pair production
at the Tevatron calculated for various $M_{W'}$ and $g'$.
The data, predictions from MCFM, and contributions needed from new physics
are listed in the last three rows.
\label{table1}
}
\medskip
\begin{ruledtabular}
\begin{tabular}{cc|cc|cc|cc}
$M_{W'}$ (GeV) & $g'$ & $\sigma_{t\bar t}$ (pb) & $A^{t\bar t}$ & 
\multicolumn{2}{c|}{$A^{t\bar t}$} & \multicolumn{2}{c}{$A^{t\bar t}$}  \\
              &      &                        &              & 
$|\Delta y | < 1$ & $|\Delta y | > 1$ & $M_{t\bar t} < 450$ GeV & 
  $M_{t\bar t} > 450$ GeV \\
\hline
\hline
$200$ & $0.85$ & $7.99$ & $0.129$ & $0.044$ & $0.321$ & $0.061$ & $0.217$ \\
$300$ & $1.2$  & $8.28$ & $0.151$ & $0.065$ & $0.348$ & $0.062$ & $0.257$ \\
$400$ & $1.5$  & $8.24$ & $0.140$ & $0.063$ & $0.324$ & $0.050$ & $0.247$ \\
$500$ & $1.8$  & $8.21$ & $0.132$ & $0.060$ & $0.305$ & $0.042$ & $0.237$ \\
$600$ & $2.1$  & $8.19$ & $0.125$ & $0.058$ & $0.290$ & $0.036$ & $0.229$ \\
\hline
\multicolumn{2}{c|}{Data (parton)} & {\scriptsize $7.70\pm 0.52$} & 
{\scriptsize $0.158\pm0.074$} & {\scriptsize $0.026 \pm0.118$} & 
{\scriptsize $0.611\pm0.256$} & {\scriptsize $-0.116\pm0.153$} &
{\scriptsize $0.475\pm0.112$} \\
\multicolumn{2}{c|}{MCFM} & {\scriptsize $7.45^{+0.72}_{-0.63}$} & 
{\scriptsize $0.058\pm0.009$} & {\scriptsize $0.039 \pm0.006$} & 
{\scriptsize $0.123\pm0.018$} & {\scriptsize $0.04\pm0.006$} &
{\scriptsize $0.088\pm0.0013$} \\
\multicolumn{2}{c|}{New Physics} & -- & 
{\scriptsize $0.100\pm0.074$} &  -- & 
{\scriptsize $0.488\pm0.257$} & -- &
{\scriptsize $0.387\pm0.112$} \\
\end{tabular}
\end{ruledtabular}
\end{table}

\subsection{Fit to the data}
The data \cite{cdf1}, the predictions from MCFM \cite{mcfm}, 
and the contributions from the new physics needed to explain the data
are summarized in the last three rows of Table~\ref{table1}.  
The entries for the total cross section, $A^{t\bar t} (|\Delta y|<1)$,
and $A^{t\bar t} (M_{t\bar t} < 450\;{\rm GeV})$ are consistent between
the data and the MCFM, so that no contributions are needed from new physics,
as indicated by ``--'' in the last row.
The deviations for $A^{t\bar t} (|\Delta y|>1)$ and 
$A^{t\bar t} (M_{t\bar t} > 450\;{\rm GeV})$ are about $2\sigma$ and
$3.5\sigma$, respectively.
In Table~\ref{table1}, we show the results for $M_{W'}= 200 -600$ GeV
with appropriate $g'$s.  They all give consistent total cross sections
with the $\sigma_{t\bar t}$ \cite{cdf2} within $1\sigma$. Also, it was
shown in our previous work \cite{ours} that the choices are consistent
with the invariant mass $M_{t\bar t}$ distribution \cite{cdf-mtt} as well.
The predictions for low $|\Delta y|<1$ and small $M_{t\bar t} < 450$ GeV
are consistent with the data.  Most strikingly, the 
predictions for large $|\Delta y|>1$ and large $M_{t\bar t} > 450$ GeV
can be brought to be within $1\sigma$ and $1.5\sigma$, respectively,
of the difference between the data and the MCFM prediction.

\section{Direct $W^\prime$ Production at the Tevatron and LHC}

The flavor-changing $W'$ considered in this work is indeed quite light.
It can be directly produced at the Tevatron and the LHC. In the following,
we calculate the production cross sections of the $W'$ associated with
a top quark/antiquark at the Tevatron and the LHC, as well as 
compare it to the irreducible QCD background.  

\begin{figure}[t!]
\includegraphics[width=6in]{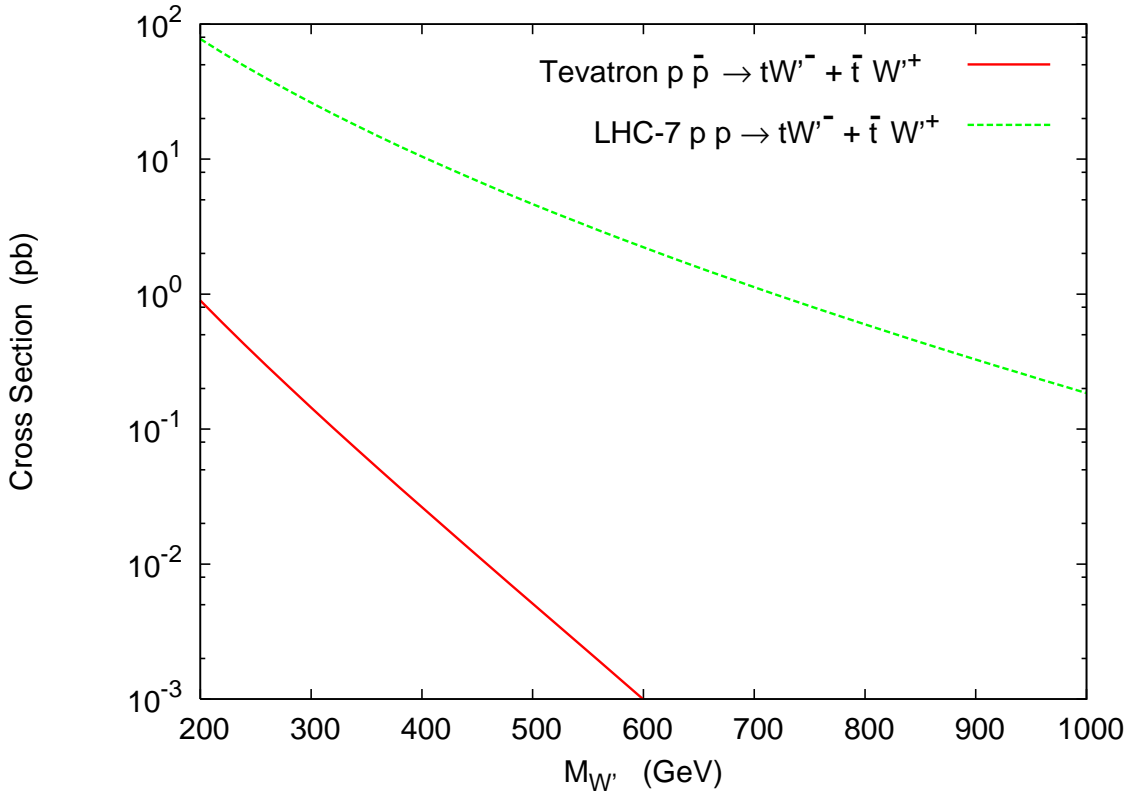}
\caption{\small \label{x-sec}
Total cross section in pb for production of
$p\bar p \to t W'^-$ and $\bar t W'^+$ at the Tevatron and
$p p \to t W'^-$ and $\bar t W'^+$ at the LHC.
}
\end{figure}

There are two Feynman diagrams for $W^\prime$ production at the hadron collider
via the subprocess $g(p_1) + d(p_2) \to t (k_1) + W^{\prime -} (k_2)$
with the $s$- and $t$-channel of down and top quark exchange, respectively.
Ignoring the $d$ quark mass, the spin- and color-summed amplitude 
squared for this process is given by
\begin{equation}
\sum \vert {\cal M} \vert^2 = 8 g^2_s g^{\prime 2} \left( g^2_L + g^2_R \right)
\left[
\frac{1}{\hat s^2} F_s + \frac{1}{\left( t - m^2_t \right)^2} F_t 
+ \frac{2}{\hat s \left( t - m_t^2 \right)} F_{st}
\right]
\end{equation}
with
\begin{eqnarray}
F_s  &=&  - \hat s 
\left[ 
\hat s + 2 t - 2 m^2_t - \frac{1}{M^2_{W'}} 
\left( \hat s - m^2_t \right) \left( \hat s + t - m_t^2 \right)
\right]  \; \; , \\
F_t & = & - \left[ m^4_t - 2 m^2_t \hat s + t \left( 2 \hat s + t \right) \right] + 4 m^2_t M^2_{W'} \nonumber \\
&  & + \frac{t}{M^2_{W'}} 
\left[ 
m^4_t - m^2_t \left( \hat s + 4 t \right) + t \left( \hat s + t \right) 
\right] \;\; , \\
F_{st} & = & \left( \hat s - m^2_t \right) \left( t - m^2_t \right)
+ \left[ m^2_t  + 2 \left( \hat s + t \right) \right] M^2_{W'} - 2 M^4_{W'} 
\nonumber \\
&& - \frac{t}{M^2_{W'}} 
\left[ 
m^4_t - 2 m^2_t \hat s + \hat s \left( \hat s + t \right)
\right] \; \; ,
\end{eqnarray}
where
$\hat s = (p_1 + p_2)^2$, $t = (p_1 - k_1)^2$ and $u = (p_1 - k_2)^2$.
The initial spin- and color-averaged amplitude squared is given by
\begin{equation}
\overline{\sum \vert {\cal M} \vert^2} = \frac{1}{2 \cdot 2} 
\frac{1}{3 \cdot 8} 
\sum \vert {\cal M} \vert^2 \; .
\end{equation}
The differential cross section versus the angle $\theta$ 
(the angle between the momenta of outgoing top and the incoming gluon) is then
\begin{equation}
\frac{d \hat \sigma}{d \cos \theta} = \frac{1}{32 \pi \hat s} 
\left( \frac{p_f}{p_i} \right) 
\overline{\sum \vert {\cal M} \vert^2} \; ,
\end{equation}
where $p_i = \sqrt{\hat s} /2$ and 
\begin{equation}
\label{pf}
p_f = \frac{1}{2 \sqrt{\hat s}} 
\left[
\hat s^2 - 2 \hat s \left( m^2_t + M^2_{W'} \right) + 
\left( m^2_t - M^2_{W'} \right)^2 \right]^{1/2} \;\; .
\end{equation}
We show in Fig.~\ref{x-sec} the total cross section for production of
$p\bar p \to t W'^-$ and $\bar t W'^+$ at the Tevatron and
$p p \to t W'^-$ and $\bar t W'^+$ at the LHC. Even if the $W'$ decays
100\% into $t \bar d$ or $\bar t d$, the size of $t\bar t$ production
cross section that it can increase is at most $0.8$ pb 
(for a 200 GeV $W'$) for the Tevatron case, 
which is about the same size as the $1\sigma$ error in the $t\bar t$ 
cross section measurement at the Tevatron.

In our scenario, the $W'^- \, (W'^+)$ decays 100\% into the 
$\bar t d \,(t \bar d)$.  
Therefore, the final state consists of a top-quark
pair plus a jet, among which one of the top quarks and the jet reconstructed
at the $W'$ mass.  The irreducible background would be QCD production
of $t\bar t + 1j$.

Recall that the top quark has a branching ratio $\sim 0.7$ decaying 
hadronically and a branching ratio $\sim 0.22$ decaying semi-leptonically
(only counting the $e,\mu$ modes).
We require the top quark that comes from the $W'$ decay
decays hadronically, in order to have a fully reconstructed top quark.
On the other hand, we require the other top to decay semi-leptonically, in order
to have a cleaner jet combinations in the final state. 
We adopt a simple parton-level analysis with the energy-momentum of the
jets and leptons smeared by
\[
\frac{\Delta E}{E} = \frac{1.0}{\sqrt{E}} \oplus 0.02 \;.
\]
We impose the following kinematic cuts for detection of the leptons and the jets
\begin{equation}
\label{tev-cut}
{\rm Tevatron} : 
\begin{cases} 
\; p_{T_{\ell}} > 15 \;{\rm GeV}   \;, & | \eta_\ell | < 2 \\
\; p_{T_{j}} > 15 \;{\rm GeV}   \;, & | \eta_j | < 2   
\end{cases}
\end{equation} 
at the Tevatron.  The kinematic cuts for the LHC are
\begin{equation}
\label{lhc-cut}
{\rm LHC} : 
\begin{cases} 
\; p_{T_{\ell}} > 20 \;{\rm GeV}   \;, & | \eta_\ell | < 2.5 \\
\; p_{T_{j}} > 20 \;{\rm GeV}   \;, & | \eta_j | < 2.5   
\end{cases}
\end{equation}

\begin{figure}[t!]
\centering
\includegraphics[width=3.2in]{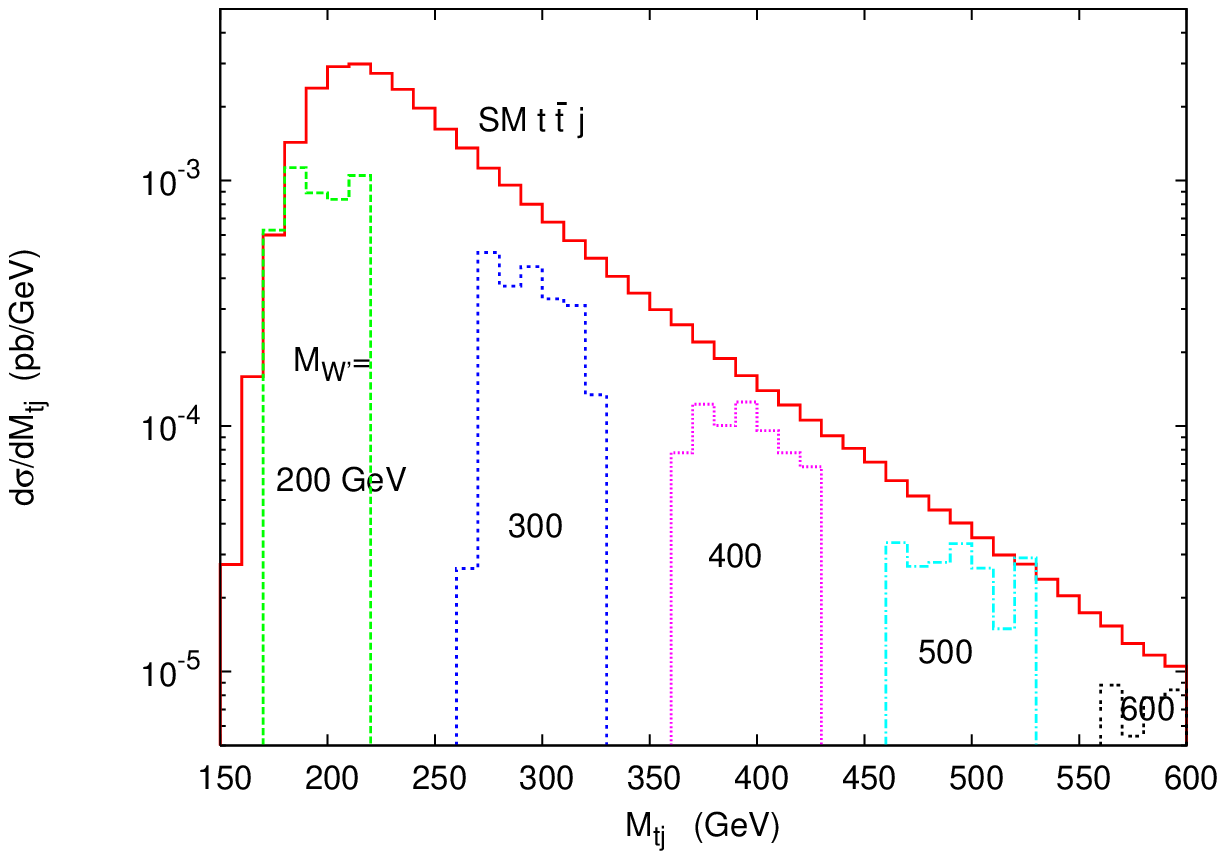}
\includegraphics[width=3.2in]{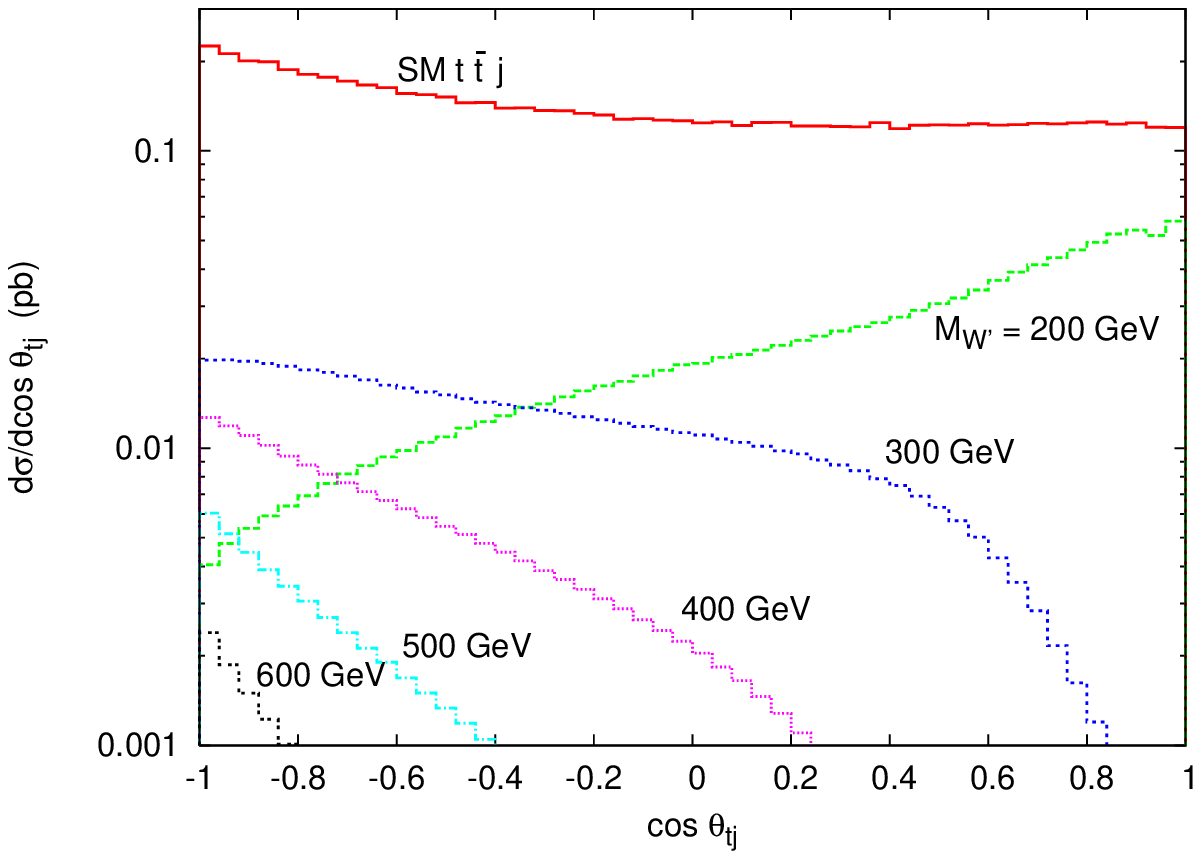}
\caption{\small \label{tev-fig}
Distributions of (a) the invariant mass and (b) cosine of the angle
between the top quark and the jet for the Tevatron. Kinematic cuts
given in Eq.(\ref{tev-cut}) have been imposed.
}
\end{figure}

\begin{figure}[t!]
\centering
\includegraphics[width=3.2in]{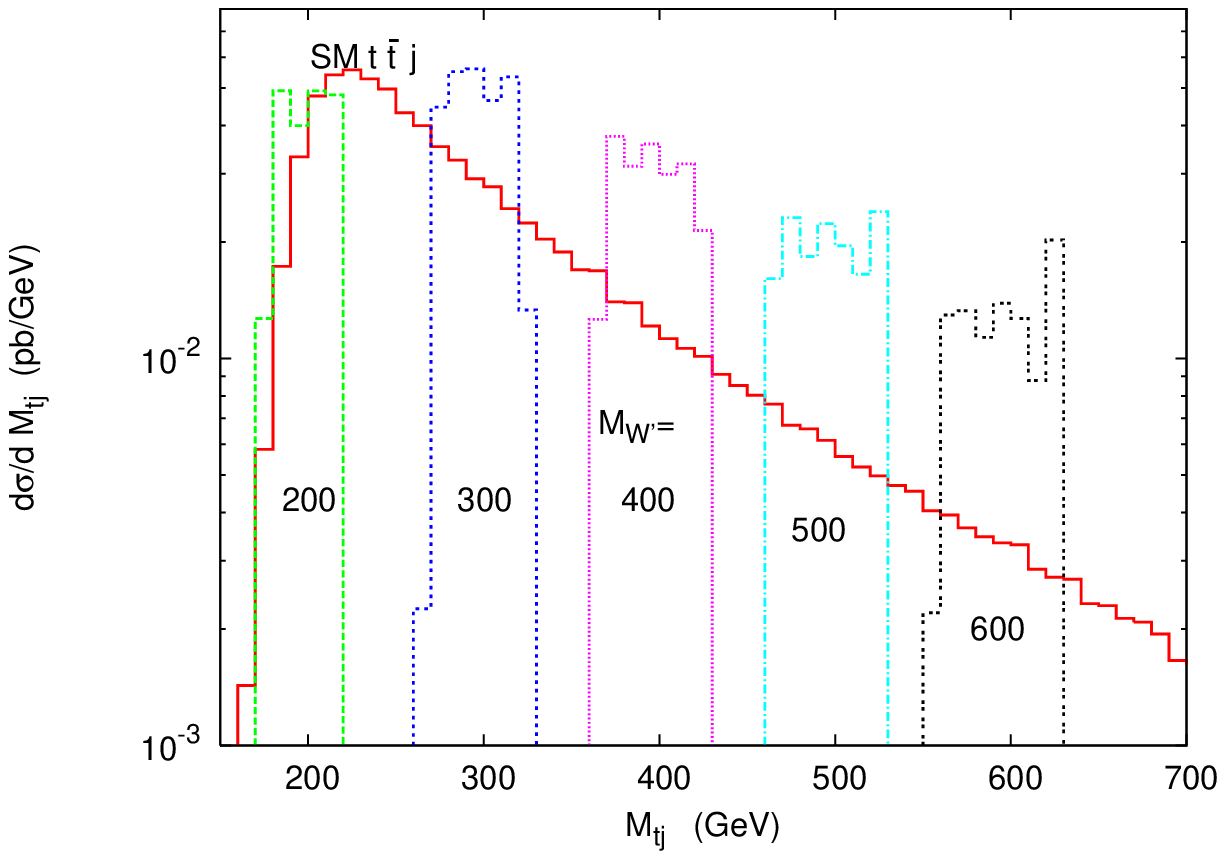}
\includegraphics[width=3.2in]{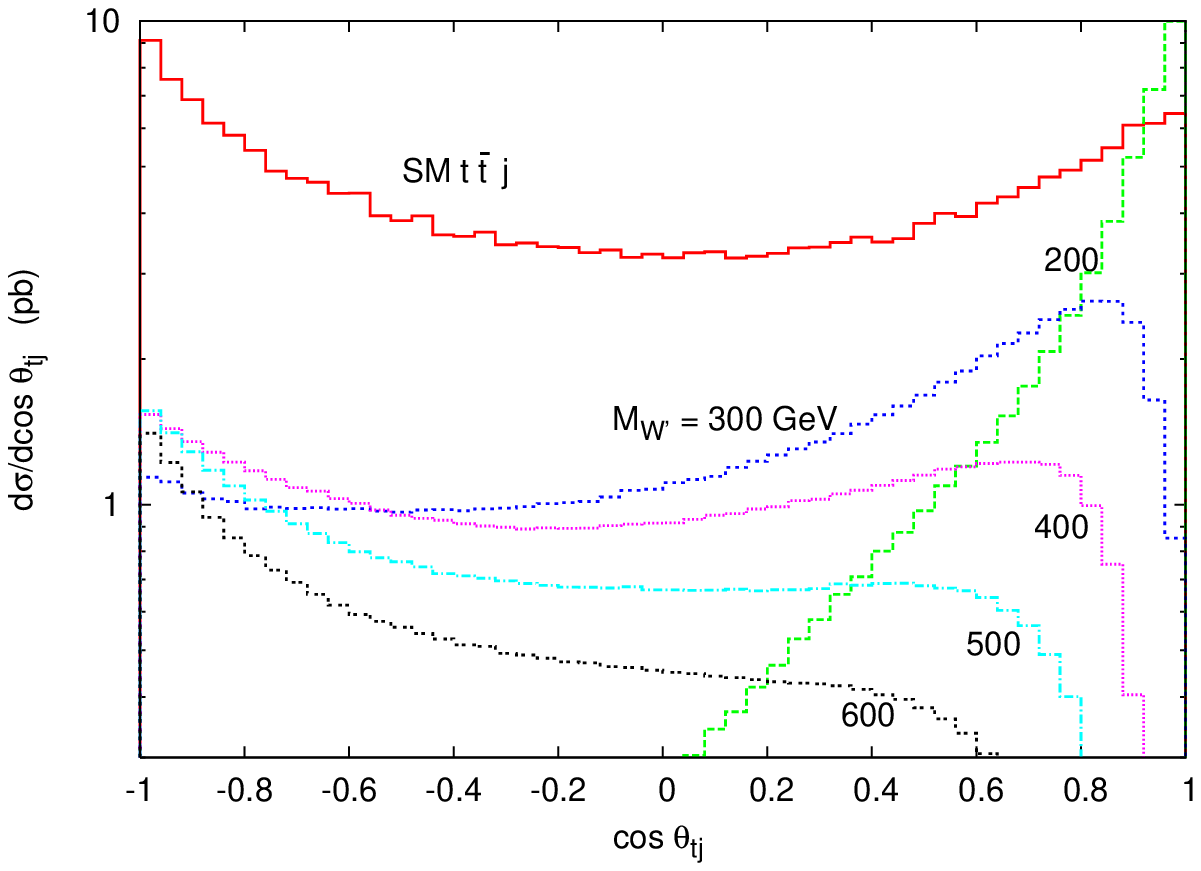}
\caption{\small \label{lhc-fig}
Distributions of (a) the invariant mass and (b) cosine of the angle
between the top quark and the jet for the LHC-7 (7 TeV). Kinematic cuts
given in Eq.(\ref{lhc-cut}) have been imposed.
}
\end{figure}

We anticipate the most distinguishable
 distributions between the signal
and background are the invariant mass $M_{tj}$ and the cosine of
the angle between the top quark (coming from the $W'$ or the hadronic
top in the background) and the jet.  
These distributions can show the difference between the signal
and the background  mainly due to the decay from the $W'$ in the signal.
On the other hand, the jet in the background most of the time 
radiates off an initial quark or gluon leg.  Thus, there is no particular 
separable angle between the hadronic top and the quark, as well as
a specific invariant mass for the $(t,j)$. 
We show these distributions
for the Tevatron in Fig.~\ref{tev-fig} and for the LHC-7 (7 TeV)
in Fig.~\ref{lhc-fig}.  For each $W'$ of mass $M_{W'}$ we use 
the value of $g'$ given in Table~\ref{table1}.
It is clear that the $M_{tj}$ for the background is a continuum while
that of the signal peaks around the $W'$ mass. Also, the $\cos \theta_{tj}$
shows that when the $W'$ is light ($\sim 200$ GeV) the opening angle
between the top and the jet tends to be quite narrow, but this feature
is lost when $W'$ becomes heavier.  

We perform an event counting for both the signal and background at the
Tevatron and the LHC.  For example, if we are searching for a 200 GeV $W'$
we will look at the $M_{tj}$ distribution and count the number of events
under the range $200 \pm \Delta$ GeV.  As indicated in Fig.~\ref{tev-fig} 
the spread of the resonance peak is about 10\% of the $W'$ mass at 
the Tevatron, we choose $\Delta = 0.1 M_{W'}$.  That is, we look 
under $200 \pm 20,\;300 \pm 300,\; 400 \pm 40,\; 500 \pm 50$ GeV 
for searches of $200 - 500$ GeV $W'$ resonances. In addition, we impose 
a cut of $\cos \theta_{tj} > 0$ for the search of 200 GeV $W'$ but not 
the others.  We show the number of events for the signal and background
at the Tevatron for an integrated luminosity of $10$ fb$^{-1}$ 
in Table~\ref{tev-no}.  
We repeat the same exercise for the LHC choosing the same 
$\Delta = 0.1 M_{W'}$, and show the number of events 
with an integrated luminosity of $0.1$ and $1$ fb$^{-1}$ in Table~\ref{lhc-no}.

\begin{table}[b!]
\caption{\small \label{tev-no}
The number of events for the $W'$ signal and the background under the
distribution $ 0.9 M_{W'} < M_{tj} < 1.1 M_{W'}$ with an integrated luminosity
of 10 fb$^{-1}$ at the Tevatron. An additional cut of $\cos\theta_{tj} >0$
for the 200 GeV $W'$ only. }
\medskip
\begin{ruledtabular}
\begin{tabular}{cc|cc|cc}
$M_{W'}$ (GeV) & $g'$ & No. of signal events $S$ 
  & No. of background events $B$ & $S/B$ & $S/\sqrt{B}$ \\
\hline
$200$ & $0.85$ & $285$ & $640$ & $0.44$ & $11$  \\   
$300$ & $1.2$  & $210$ & $460$ & $0.46$ & $9.8$   \\
$400$ & $1.5$  & $67$  & $130$ & $0.52$ & $5.9$  \\
$500$ & $1.8$  & $19$  & $40$  & $0.48$ & $3.0$  \\
$600$ & $2.1$  & $5$   & $14$  & $0.36$ & $1.3$  
\end{tabular}
\end{ruledtabular}
\end{table}

\begin{table}[b!]
\caption{\small \label{lhc-no}
The number of events for the $W'$ signal and the background under the
distribution $ 0.9 M_{W'} < M_{tj} < 1.1 M_{W'}$ with an integrated luminosity
of $0.1\; (1)$ fb$^{-1}$ at the LHC. An additional cut of $\cos\theta_{tj} >0$
for the 200 GeV $W'$ only. }
\medskip
\begin{ruledtabular}
\begin{tabular}{cc|cc|cc}
$M_{W'}$ (GeV) & $g'$ & No. of signal events $S$ 
  & No. of background events $B$ & $S/B$ & $S/\sqrt{B}$ \\
\hline
$200$ & $0.85$ & $180\;(1800)$ & $130\;(1300)$& $1.4$ & $16\;(50)$  \\   
$300$ & $1.2$  & $270\;(2700)$ & $170\;(1700)$& $1.6$ & $21\;(65)$  \\
$400$ & $1.5$  & $200\;(2000)$ & $98\;(980)$  & $2.0$ & $20\;(64)$  \\
$500$ & $1.8$  & $140\;(1400)$ & $60\;(600)$  & $2.3$ & $18\;(57)$  \\
$600$ & $2.1$  & $96\;(960)$   & $39\;(390)$  & $2.4$ & $15\;(49)$  
\end{tabular}
\end{ruledtabular}
\end{table}

In Table~\ref{tev-no}, the value of $g'$ used for each $M_{W'}$ is 
according to what has been used to explain the top FBA. The ratio of
$S/B$ is about $0.5$ for all $M_{W'}$ but the significance $S/\sqrt{B}$
ranges from about 11 to 1 for $M_{W'} = 200 - 600$ GeV.
It is implied from Table~\ref{tev-no} Tevatron would have a good chance
observing the $W'$ up to about 400 GeV that could be the explanation
for the top FBA.
Furthermore, the observability improves substantially at the LHC-7.
The chance of observing the $W'$ all the way to 600 GeV is very 
promising at the LHC-7 with just 100 pb$^{-1}$ luminosity, as shown
by the significance $S/\sqrt{B} = 15 -21$ in Table~\ref{lhc-no}.  
Further improvement by increasing the luminosity to 1 fb$^{-1}$ 
can push the significance to more than 50 at the LHC.

Similar analysis at the LHC can be found in Refs.~\cite{kim} 
and \cite{Barger:2011ih}.
There was another work using charge asymmetry at the LHC to probe
the $W'$ boson \cite{strassler}. 

\section{Conclusions}

We have shown that with a new particle exchanged in $t$-channel the 
forward-backward asymmetry in $t\bar t$ production increases with
the invariant mass $M_{t\bar t}$ and the rapidity difference $|\Delta y|$.
This is a generic feature for any particle exchanged in $t$-channel. 
We have also demonstrated that the new CDF data on FBA can be accommodated 
using a flavor-changing pure right-handed $W'$ boson, which couples 
only to the $d$ and $t$ quarks with an appropriate coupling constant
$g'$.  We can bring the FBA to within $1.5\sigma$ of the data
in the large $M_{t\bar t} > 450$ GeV region and within $1\sigma$ 
in the large rapidity difference $|\Delta y|$ region.  The specific $W'$
model that we proposed is consistent with existing data on the direct
search and with flavor-changing current data.
Some attempts to find a realistic model for such a 
flavor-changing gauge boson were in Ref.~\cite{flavor}.

Furthermore, we have shown that such a $W'$ up to about 400 GeV 
is readily observed at the Tevatron with an integrated luminosity of 
10 fb$^{-1}$, and at the LHC with an integrated luminosity of 
100 pb$^{-1}$, which should be within the current year of running.
The signal-background analysis that we performed is based on
parton-level calculations. More realistic simulation may be necessary.
Nevertheless, the present work has indicated that the $W'$ really has
a good chance to be seen. 
The cleanest signal of the $W'$
would be the sharp peak in the invariant mass $M_{tj}$ distribution. 
By counting the number of events below the peak for the signal and
background, the significance of the signal can reach a level of 10
at the Tevatron and a level of 60 at the LHC.

\section*{\bf Acknowledgments}

This research was supported in parts by the NSC under Grant
Nos. 99-2112-M-007-005-MY3 and 98-2112-M-001-014-MY3, by the NCTS, and
by WCU program through the NRF funded by the MEST
(R31-2008-000-10057-0).


\end{document}